\documentclass[reprint, superscriptaddress, amsmath, amssymb, notitlepage, nofootinbib, 10pt, aps]{revtex4-1}

\usepackage{color}
\usepackage{graphicx}
\usepackage{bmpsize}
\usepackage{dcolumn}

\usepackage[T1]{fontenc}

\usepackage{bm}
\usepackage{hyperref}
\usepackage{cleveref}
\usepackage{braket}
\usepackage{mathtools}

\begin{document}

\preprint{APS/123-QED}

\title{Adiabatic transport in one-dimensional systems with a single defect}

\author{Kazuaki Takasan}
\affiliation{
Department of Physics, University of California, Berkeley, California 94720, USA}
\affiliation{
Materials Sciences Division, Lawrence Berkeley National Laboratory, Berkeley, CA 94720, USA
}

\author{Masaki Oshikawa}
\affiliation{
Institute for Solid State Physics, University of Tokyo, Kashiwa, Chiba 277-8581, Japan}
\affiliation{
Kavli Institute for the Physics and Mathematics of the Universe (WPI), University of Tokyo, Kashiwa, Chiba 277-8583, Japan}

\author{Haruki Watanabe}
\email{hwatanabe@g.ecc.u-tokyo.ac.jp}
\affiliation{
Department of Applied Physics, University of Tokyo, Tokyo 113-8656, Japan}

\date{\today}

\begin{abstract}
The adiabatic transport properties of U(1) invariant systems are determined by the dependence of the ground state energy on the twisted boundary condition. We examine a one-dimensional tight-binding model in the presence of a single defect and find that the ground state energy of the model shows a universal dependence on the twist angle that can be fully characterized by the transmission coefficient of the scattering by the defect. 
We identify resulting pathological behaviors of Drude weights in the large system size limit: (i) both the linear and nonlinear Drude weights depend on the twist angle and (ii) the $N$-th order Drude weight diverges proportionally to the $(N-1)$-th power of the system size. 
To clarify the physical implication of the divergence, we simulate the real-time dynamics of the tight-binding model under a static electric field and show that the divergence does not necessarily imply the large current. Furthermore, we address the relation between our results and the boundary conformal field theory.
\end{abstract}

\maketitle

\section{Introduction}
Transport phenomenon is one of the most fundamental problems in the condensed matter and statistical physics. Compared to the widely-studied linear response, the nonlinear ones have been less explored and still being an intriguing topic. For instance, rectification current~\cite{Tan2016, Tokura2018} and high-harmonic generation~\cite{Kruchinin2018, Ghimire2019} originating from nonlinear responses are extensively studied recently. Although the theoretical sides have also been actively investigated and various interesting phenomena have been proposed~\cite{Sodemann2015, Morimoto2016, deJuan2017, Dan2019, Isobe2020, Ahn2020, Takasan2020}, we still do not reach the systematic understanding of them. In particular, the general aspects of the nonlinear responses in quantum many-body systems has been less studied~\cite{Shimizu1}.

Among various transport phenomena, adiabatic transport, which is realized in the adiabatic insertion of U(1) magnetic flux, is one of the most basic ones. This is because such transport in a clean system is characterized by the Drude weight~\cite{Resta_2018} which is easily calculable from the energy eigenvalues under a twisted boundary condition using the Kohn formula~\cite{Kohn1964}. The linear Drude weight has played an essential role in the studies of transport phenomena in one-dimensional (1D) quantum many-body systems at zero and finite temperature~\cite{ShastrySutherland-twisted, Sutherland1990, Korepin1991, Castella1995, Narozhny1998, Zotos1999, Bertini2021}. Also, the adiabatic current induced in a small metallic ring pierced by magnetic flux is known as persistent current. This is a manifestation of the Aharanov-Bohm effect in quantum many-body systems and has been well-studied both experimentally and theoretically~\cite{ImryBook, Viefers2004, Kulik2010, Bleszynski-Jayich2009, Bluhm2009, Cheung1988, Loss1992, Gogolin1994, Meden2003, Dias2006}. Note that the Drude weight corresponds to a flux derivative of the persistent current, as explained later.

Recently, nonlinear Drude weights were introduced as a direct generalization of the Drude weight of linear response to higher-order responses~\cite{PhysRevB.102.165137}. The generalized Kohn formula is also derived~\cite{PhysRevB.102.165137} and allows us to evaluate the nonlinear Drude weights easily. For non-interacting band insulators, the Drude weights underlie the phenomenon so-called Bloch oscillation~\cite{PhysRevB.102.165137}. However, it turns out that nonlinear Drude weights in the spin-1/2 XXZ Heisenberg chain diverges in the limit of large system size, depending on the order of the response and the value of the anisotropy parameter~\cite{PhysRevB.102.165137,2103.05838}. 
While the origin of the divergence in the XXZ chain was discussed in Ref.~\cite{2103.05838}, the general condition, such as if the many-body interaction is needed or not for the divergence, is still unknown. It is also unclear what happens under a static electric field in the system with the divergent Drude weights. It is important to clarify the above problems in order to reach the systematic understanding of the nonlinear responses.

In this work, we examine the dependence of the ground state energy to the twisted boundary condition in general 1D systems and clarify the relation to the adiabatic transport including the nonlinear responses. We then establish a concrete example in which nonlinear Drude weights diverge even \emph{without} many-body interactions. Our model is a single-band tight-binding model with a single defect. Regardless of the details of the defect, the divergence of nonlinear Drude weights is much stronger than those observed in the XXZ model. Furthermore, we find that even the linear Drude weight shows a pathological behavior, depending strongly on the twisted boundary condition.

We also elucidate the physical consequence of the diverging behavior of the Drude weights. One might think that the the divergence implies an arbitrary large current response when the system is under a static electric field. To clarify this point, we employ the numerical real-time simulation and show that the adiabatic transport is rather suppressed compared to the defect-free case. The real-time simulations have not been done to study the divergent Drude weights because the interaction makes the long-time simulation difficult. In contrast, our model is non-interacting and easy to reach the adiabatic limit where the Drude weights dominate. This is one of the advantageous points of our single-defect model.

We emphasize that in this paper we discuss the Drude weights which characterize the adiabatic response of the uniform current to the applied uniform electric field. In other words, we focus on the Drude weights  determined by (generalized) Kohn formulas~\cite{Kohn1964, PhysRevB.102.165137}. They are sensitive to the presence of the defect, or equivalently to the boundary condition. In fact, it follows trivially that those Drude weights exactly vanish for the open boundary condition (the limit of an infinitely strong defect), since the uniform component of the current is zero. One may expect that the conduction property of the bulk should be independent of the boundary condition. In fact, it has been pointed out that the Drude weight for the periodic boundary condition can be extracted even under the open boundary condition with an appropriate procedure~\cite{RigolShastry-Drude,BellomiaResta}. However, it is the purpose of the present paper to elucidate the effects of the single defect on the adiabatic transport, rather than the bulk conduction properties. We hope that the insights gained in the present study will be also useful in understanding the bulk conduction properties of more general quantum many-body systems.

This article is organized as follows: In Sec.~\ref{SecII}, we define the several quantities characterizing the adiabatic transport in the large size limit and calculate the quantity in several well-known models as examples. In Sec.~\ref{SecIII}, we introduce our central model in this study, a single-band tight-binding model with a defect, and analytically show that the model has the pathological properties including the divergence of the Drude weights. In Sec.~\ref{SecIV}, we study the physical implication of the divergence using the numerical real-time simulation. Finally, we mention the connection to the boundary conformal field theory (CFT) in Sec.~\ref{SecV} and then summarize this paper in Sec.~\ref{SecVI}.

\section{Twisted boundary condition and the ground state energy}
\label{SecII}
Before discussing our concrete model, here we summarize the relation between the dependence of the ground state energy on the twisted boundary condition and the adiabatic transport of the system.
\subsection{General consideration}
\label{general}
Let us consider 1D systems described by a local Hamiltonian $\hat{H}$ that contains only finite-ranged hoppings and finite-ranged interactions. We impose the periodic boundary condition with the system size $L$. Assuming a global U(1) symmetry for the particle number conservation, we introduce a uniform vector potential $A=\theta/L$. 

We are interested in the $\theta$-dependence of the ground state energy $E_{\text{GS}}^L(\theta)$ as it determines the ballistic transport property of the system.  For example, the spontaneous current density, averaged over the entire system, is given by
\begin{align}
j_{\text{GS}}^L(\theta)=\frac{dE_{\text{GS}}^L(\theta)}{d \theta}.
\label{jGS}
\end{align}
Higher-derivatives are related to the Drude weights~~\cite{Kohn1964,PhysRevB.102.165137}:
\begin{align}
\mathcal{D}_{N}^L(\theta)\equiv L^N\frac{d^{N+1}}{d\theta^{N+1}}E_{\text{GS}}^L(\theta).
\label{D_N}
\end{align}
Here $N$ represents the order of the response: $N=1$ is the linear response and $N\geq 2$ is a higher-order response.
Since $e^{i\theta}$ may be interpreted as the phase of twisted boundary condition, $E_{\text{GS}}^L(\theta)$ must have the period $2\pi$ as a function of $\theta$.  

The ground state energy can, in general, be expanded into a power-series of $L$:
\begin{align}
E_{\text{GS}}^L(\theta)&=c_{+1}(\theta) L+c_0(\theta)+c_{-1}(\theta) L^{-1}+o(L^{-1}),
\label{GSEL}
\end{align}
where, $o(L^{-n})$ represents corrections that decay faster than $L^{-n}$ in the large $L$ limit, which includes possible terms with non-integer power $\alpha\notin\mathbb{Z}$ and $\alpha>1$. By definition, coefficients $c_p(\theta)$ ($p=1,0,-1,\cdots$) do not depend on $L$. Note that $c_{+1}(\theta)$ corresponds to the energy density $\varepsilon_{\text{GS}}$ in the thermodynamic limit and cannot depend on $\theta$. 

In order to investigate the $\theta$-dependence of $c_0(\theta)$ and $c_{-1}(\theta)$, let us derive a bound for the spontaneous current density $j_{\text{GS}}^L(\theta)$ following the proof of the Bloch theorem~\cite{PhysRev.75.502,WatanabeBloch}. To this end, we introduce the large gauge transformation operator
\begin{equation}
\hat{U}\equiv e^{i(2\pi m/L)\sum_{x}x\hat{n}_x}
\end{equation}
that changes the flux $\theta$ by $2\pi m$, hence the gauge field $A=\theta/L$ by $2\pi m/L$. Here $\hat{n}_x$ is the number density operator at the site $x$.
Therefore,
\begin{equation}
\hat{U}^\dagger \hat{H} \hat{U}-\hat{H}=\frac{2\pi m}{L}\frac{d\hat{H}}{d A}+\frac{1}{2}\left(\frac{2\pi m}{L}\right)^2\frac{d^2\hat{H}}{d A^2}+O(L^{-2}).
\end{equation}
Here, $O(L^{-n})$ is a quantity that decays either equally fast with or faster than $L^{-n}$.
We take the expectation value of this relation using the ground state of $\hat{H}$. Since the variational principle implies $\langle\hat{U}^\dagger \hat{H} \hat{U}-\hat{H}\rangle\geq0$, we find
\begin{equation}
mj_{\text{GS}}^L(\theta)+\pi m^2\Big\langle\frac{d^2\hat{H}}{d \theta^2}\Big\rangle+O(L^{-2})\geq0
\label{mBloch}
\end{equation}
for any $m\in\mathbb{Z}$. It follows that $\langle d^2\hat{H}/d \theta^2\rangle>0$ and that $c_{0}(\theta)$ is  $\theta$-independent. In fact, a non-zero $c_0(\theta)$ indicates the energy due to a defect, and thus $c_0(\theta)$ generally vanishes in translation-invariant systems. The statement $c_0(\theta)=0$ also follows from the conformal mapping from the infinite plane to a cylinder~\cite{BCN1986,Affleck-FSSCFT}, when the translation-invariant system is described by a CFT.

On the other hand, $c_{-1}(\theta)$ is related to the central charge and conformal dimensions of the CFT, and thus is expected to be universal (see the later sections for its precise meaning).
Non-vanishing $\theta$-dependence starts at $c_{-1}(\theta)$, indicating the universal nature of the transport we will discuss in this paper.
To quantify the $\theta$-dependence, let us look at the $n$-th derivative:
\begin{align}
d_{n}(\theta)\equiv\frac{d^{n}c_{-1}(\theta)}{d\theta^{n}}.
\label{dN}
\end{align}
The inequality \eqref{mBloch} implies
\begin{equation}
\left|d_{1}(\theta)\right|\leq\pi\lim_{L\to\infty}L\Big\langle\frac{d^2\hat{H}}{d\theta^2}\Big\rangle=\lim_{L\to\infty}\int_{-\infty}^{\infty}d\omega \sigma^L(\omega).\label{newbound}
\end{equation}
The last equality is the frequency sum rule of the optical conductivity $\sigma^L(\omega)$. Furthermore, the positivity of the regular part of $\text{Re}[\sigma^L(\omega)]$ implies:
\begin{equation}
d_2(\theta)\leq \lim_{L\to\infty}L \Big\langle\frac{d^2\hat{H}}{d\theta^2}\Big\rangle
=\frac{1}{\pi}\lim_{L\to\infty}\int_{-\infty}^{\infty}d\omega \sigma^L(\omega).
\label{drudebound}
\end{equation}
See Appendix~\ref{optical} for the derivation. These inequalities give general bounds for the spontaneous current density $j_{\text{GS}}^L(\theta)=d_1(\theta)L^{-1}+o(L^{-1})$ and the linear Drude weight $\mathcal{D}_1^L(\theta)=d_2(\theta)+o(1)$.

By definition, the Drude weights are represented as $\mathcal{D}_{N}^L(\theta)=d_{N+1}(\theta)L^{N-1} + o(L^{N-1})$ in general. When $d_{N+1}(\theta)\neq0$ for an $N\geq2$, two important consequences follow immediately: (i) The $N$-th order Drude weight $\mathcal{D}_{N}^L(\theta)$ ($N\geq2$) diverges in the large $L$ limit with the power $L^{N-1}$, and (ii) The \emph{linear} Drude weight $\mathcal{D}_{1}^L(\theta)$ depends non-trivially on $\theta$ even in the large $L$ limit.  Since a physical quantity in a sufficiently large system must be insensitive to the choice of the boundary condition, this implies that even the linear Drude weight becomes unphysical as a bulk quantity when $d_{n}(\theta)\neq0$ for an $n\geq3$.

Conversely, this discussion suggests that $d_{n}(\theta)$ vanishes for all $n\geq3$ whenever the linear Drude weight is supposed to be a good physical quantity independent of $\theta$. This is expected to be the case when the system has both U(1) symmetry and the lattice translation symmetry, regardless of the presence or the absence of many-body interactions. Even in this case the nonlinear Drude weight $\mathcal{D}_{N}^L(\theta)$ may still diverge but with a power smaller than $L^{N-1}$. This is the case of the XXZ model as we discuss later.

\subsection{Single-band tight-binding model}
\label{secH0}
As an example, let us consider a single-band tight-binding model of spinless electrons in one dimension. 
\begin{align}
\hat{H}_0&\equiv-t_0\sum_{x=1-L/2}^{L/2}(\hat{c}_{x+1}^\dagger e^{-i\theta/L}\hat{c}_x+\text{h.c.}).
\label{H0}
\end{align}
The lattice constant is set $1$. We assume that the system size is $L=4\ell'+2$ and the number of electrons is $N_{\text{el}}=2\ell+1$ ($\ell, \ell'\in\mathbb{N}$). The Hamiltonian can be diagonalized by the Fourier transformation $\hat{c}_{k_n}^\dagger\equiv L^{-1/2}\sum_{x}\hat{c}_{x}^\dagger e^{ik_n x}$. We get $\hat{H}_0=\sum_{n=1}^{L}\epsilon_{k_n}\hat{c}_{k_n}^\dagger\hat{c}_{k_n}$ with $k_n\equiv 2\pi n/L$ and $\epsilon_{k_n}=-2t_0\cos(k_n+\theta/L)$.  For the range $|\theta|<\pi$, the ground-state energy is given by
\begin{align}
E_{\text{GS}}^L(\theta)&=\sum_{n=-\ell}^{\ell}\epsilon_{k_n}=-\frac{v_F}{\sin(\pi/L)}\cos(\theta/L).
\end{align}
Here the Fermi velocity $v_F$ is defined by
\begin{equation}
v_F\equiv 2t_0\sin k_F,\quad k_F\equiv\pi N_{\text{el}}/L.\label{vF}
\end{equation}
Therefore,
\begin{align}
c_{+1}(\theta)&=-\frac{v_F}{\pi}\label{cp10},\\
c_0(\theta)&=0,\label{c010}\\
c_{-1}(\theta)&=\frac{v_F}{2\pi}[\arccos(\cos\theta)]^2-\frac{\pi v_F}{6}.\label{cm10}
\end{align}
This expression of $c_{-1}(\theta)$ respects the period $2\pi$ of $E_{\text{GS}}^L(\theta)$ and is valid for any $\theta\in\mathbb{R}$. 
Our convention of arccosine is the standard one, satisfying $0\leq\arccos(x)\leq\pi$, $\arccos(-1)=\pi$, and $\arccos(+1)=0$.  Thus $\arccos(\cos\theta)=|\theta|$ for $|\theta|<\pi$. $c_{-1}(\theta)$ and its derivative $d_{n}(\theta)$ ($n=1, 2, 3$) of this model are shown in Fig.~\ref{fig1}.

The $N$-th order Drude weight ($n\geq1$) in this model remains finite in the thermodynamic limits, which is given by  $\lim_{L\to\infty}\mathcal{D}_N^L(\theta)=(-1)^{N-1}v_F/\pi$.  More generally, all nonlinear Drude weights remain finite if the ground state energy takes the form
\begin{align}
E_{\text{GS}}^L(\theta)&=L\varepsilon_{\text{GS}}^L(\theta/L)
\end{align}
with a function $\varepsilon_{\text{GS}}^L(A)$ whose large $L$ limit $\varepsilon_{\text{GS}}(A)$ is a smooth function of $A$. In the above tight-binding model, $\varepsilon_{\text{GS}}(A)=-(v_F/\pi)\cos A$. In such a case, $c_{p}(\theta)$ ($p=1,0,-1,\cdots$) is a polynomial of $\theta$ with the maximal power $\theta^{1-p}$ and $\lim_{L\to\infty}\mathcal{D}_{N}^L(\theta)=d^{N+1}\varepsilon(A)/dA^{N+1}|_{A=0}$. 

\subsection{Tomonaga-Luttinger liquids}
\label{XXZ}
For Tomonaga-Luttinger liquids (TLLs) with the Luttinger parameter $K$ and the velocity parameter $v$, the finite-size scaling of the ground state energy is known to be~\cite{Giamarchi2004B}
\begin{align}
c_{+1}(\theta)&=\varepsilon_{\text{GS}},\\
c_0(\theta)&=0,\\
c_{-1}(\theta)&=                              
\frac{K v}{2\pi} [\arccos(\cos\theta)]^2
- \frac{\pi v}{6}.
\label{cmTLL}
\end{align}
For example, in the case of the spin-1/2 XXZ Heisenberg spin chain
\begin{equation}
\hat{H}=J\sum_{x=1}^L\left(\frac{1}{2}\hat{s}_{x+1}^+e^{-i\theta/L}\hat{s}_x^-+\text{h.c.}+\Delta\hat{s}_{x+1}^z\hat{s}_x^z\right)
\end{equation} with $-1<\Delta\leq1$, 
the parameters $\varepsilon_{\text{GS}}$, $K$, and $v$ are given by~\cite{Hamer_1987,PhysRevLett.65.1833,doi:10.1142/S0217979212440092}
\begin{align}
\varepsilon_{\text{GS}}&=\frac{J}{4}\cos\gamma-\frac{J}{2}\sin\gamma\int_{-\infty}^\infty dx\frac{\sinh[(\pi-\gamma) x]}{\sinh(\pi x)\cosh(\gamma x)},\\
v &= J\frac{\pi\sin{\gamma}}{2\gamma},\quad K =\frac{\pi}{2(\pi-\gamma)},
\end{align}
where $\gamma\equiv\arccos\Delta$. 
This model with $J=2t_0$ and $\Delta=0$ can be exactly mapped to the above tight-binding model at the half filling $N_{\text{el}}=L/2$.
$c_{-1}(\theta)$ and its derivative $d_{n}(\theta)$ ($n=1, 2, 3$) of this model are shown in Fig.~\ref{fig1}. 

Although TLLs satisfy $d_{n}=0$ for $n\geq3$, 
nonlinear Drude weights diverge in the thermodynamic limit
in the XXZ chain as studied in Refs.~\cite{PhysRevB.102.165137, 2103.05838}. This is due to the irrelevant perturbation to the CFT and not captured in the unperturbed TLLs~\cite{2103.05838}. We emphasize that this fact does not contradict to our statement because the divergence of the $N$-th order Drude weight in the XXZ chain is with a power smaller than $L^{N-1}$.

We observe that $c_{-1}(\theta)$ in Eq.~\eqref{cmTLL} is written completely in terms of the parameters of the low-energy effective theory and is universal in that sense.
In particular, the constant term $-\pi v/6$ follows from the central charge $1$ of the TLL.
The Luttinger parameter $K$ only affects the coefficient of $\theta^2$ but never generates higher power terms of $\theta$ in $c_{-1}(\theta)$. Reflecting the level crossing of many-body energy levels at $\theta=(2m-1)\pi$ ($m\in\mathbb{Z}$), the slope of $c_{-1}(\theta)$ is discontinuous at these points.  The level crossing is protected by the lattice translation symmetry; that is, the two many-body energy levels crossing at these points have distinct momenta ($0$ and $2k_F$) and they cannot repel each other. This observation motivates us to investigate translation breaking perturbations.

\section{Tight-binding model with a single defect}
\label{SecIII}
Let us introduce a defect $\hat{V}$ to the tight-binding model $\hat{H}_0$ in Sec.~\ref{secH0}. The defect  induces a level repulsion at $\theta=(2m-1)\pi$ and $E_{\text{GS}}^{L}(\theta)$ becomes a smooth periodic function of $\theta$. As a consequence, the $\theta$-dependence of $c_{-1}(\theta)$ is fundamentally modified, as we shall see in Sec.~\ref{GSE}.

\subsection{Single defect scattering}
We consider a defect $\hat{V}$ localized around the site $x=0$.
Examples include a single impurity potential
\begin{equation}
\hat{V}=w\hat{c}_0^\dagger\hat{c}_0
\label{impuritypotential}
\end{equation}
and a single bond disorder
\begin{equation}
\hat{V}=-(ve^{i\delta}-t_0)\hat{c}_{1}^\dagger e^{-i\theta/L}\hat{c}_{0}+\text{h.c.},
\label{bonddisorder}
\end{equation}
but our discussion below is not restricted to these cases. The only assumptions are that $\hat{V}$ is written in terms of operators near the origin and is a bilinear of $\hat{c}_x^\dagger$ and $\hat{c}_{x'}$.  

To solve the eigenvalue equation $(\hat{H}_0+\hat{V})|k\rangle=\epsilon_k|k\rangle$, 
we postulate the following form of the wavefunction:
\begin{align}
\psi_k(x)\equiv\tilde{\psi}_k^+e^{i kx}+\tilde{\psi}_k^-e^{-i (k+2\theta/L)x}\label{pw1}
\end{align}
for $1-L/2\leq x\ll-1$ and
\begin{align}
\psi_k(x)\equiv\psi_k^+e^{i kx}+\psi_k^-e^{-i (k+2\theta/L)x}\label{pw2}
\end{align}
for $1\ll x \leq L/2$. We may assume $0<k+\theta/L<\pi$. The energy eigenvalue is still given by $\epsilon_k=-2t_0\cos(k+\theta/L)$ as in the defect-free case, but the quantization condition imposed on $k$ is modified, as we will see below.

The defect $\hat{V}$ can be characterized by the scattering matrix
\begin{align}
S_{q}\equiv
\begin{pmatrix}
T_{q}^+&R_{q}^-\\
R_{q}^+&T_{q}^-
\end{pmatrix},
\end{align}
where $T_q^\pm$ and $R_q^\pm$ are the transmission and the reflection coefficients. It maps the incoming components ($\tilde{\psi}_k^+$ and $\psi_k^-$) to the outgoing components ($\psi_k^+$ and $\tilde{\psi}_k^-$):
\begin{align}
\begin{pmatrix}
\psi_k^+\\
\tilde{\psi}_k^-
\end{pmatrix}=S_{k+\theta/L}
\begin{pmatrix}
\tilde{\psi}_k^+\\
\psi_k^-
\end{pmatrix}.
\label{condition1}
\end{align}
For example, for the single impurity potential~\eqref{impuritypotential},
\begin{align}
T_q^\pm=\frac{2t_0\sin q}{2t_0\sin q+iw},\quad
R_q^\pm=-\frac{iw}{2t_0\sin q+iw}.
\end{align}
For the single bond disorder~\eqref{bonddisorder},
\begin{align}
&T_q^\pm=e^{\pm i\delta}\frac{2t_0v\sin q}{(t_0^2+v^2)\sin q+i(t_0^2-v^2)\cos q},\\
&R_q^\pm=-e^{\pm iq}\frac{i(t_0^2-v^2)}{(t_0^2+v^2)\sin q+i(t_0^2-v^2)\cos q}.
\end{align}

The conservation of the probability current implies that the $S$-matrix is unitary for any $q$.
Furthermore, it is also constrained by symmetries of the system. For example, the time-reversal invariance implies
\begin{align}
\begin{pmatrix}
\psi_k^+\\
\tilde{\psi}_k^-
\end{pmatrix}
=
\sigma_1
\begin{pmatrix}
\tilde{\psi}_k^+\\
\psi_k^-
\end{pmatrix}^*,
\end{align}
where $\sigma_i$ ($i=1,2,3$) is the Pauli matrix.
Together with the unitarity, we find $\sigma_1 S_q^* \sigma_1 = S_q^\dagger$, which is reduced to $T_q^+ = T_q^-$~\cite{doi:10.1063/1.531762}.
Likewise, the spatial inversion about $x=x_0$ requires
\begin{align}
\begin{pmatrix}
\tilde{\psi}_k^+\\
\psi_k^-
\end{pmatrix} &= \sigma_1 e^{2ix_0q\sigma_3} \begin{pmatrix}
\tilde{\psi}_k^+\\
\psi_k^-
\end{pmatrix},\\
\begin{pmatrix}
\psi_k^+\\
\tilde{\psi}_k^-
\end{pmatrix} &= \sigma_1 e^{2ix_0q\sigma_3}\begin{pmatrix}
\psi_k^+\\
\tilde{\psi}_k^-
\end{pmatrix},
\end{align}
implying $S_q = e^{-2iqx_0\sigma_3}\sigma_1 S_q \sigma_1 e^{2iqx_0\sigma_3}$. In terms of the matrix elements, this is equivalent to
$T_q^+ =  T_q^-$ and $R_q^- =e^{-4iqx_0}  R_q^+$. For example, the impurity potential~\eqref{impuritypotential} has the site inversion symmetry ($x_0=0$) and the bond disorder~\eqref{bonddisorder} has the bond inversion symmetry ($x_0=1/2$) when $\delta=0$.
Note that our discussions below do not assume any of these symmetries.

\subsection{Quantization condition}
Now we impose the periodic boundary condition:
\begin{align}
\begin{pmatrix}
\tilde{\psi}_k^+\\
\tilde{\psi}_k^-
\end{pmatrix}=
\begin{pmatrix}
\psi_k^+e^{ikL}\\
\psi_k^-e^{-i(k+2\theta/L)L}
\end{pmatrix}.
\label{condition2}
\end{align}
We demand the existence of nonvanishing solutions to Eqs.~\eqref{condition1} and \eqref{condition2}, which leads to the quantization condition of $q\equiv k+\theta/L$. Such a condition can be most easily implemented (see Ref.~\cite{Oshikawa-IsingGap} for a related discussion on Majorana fermions) by parametrizing the scattering matrix as
\begin{align}
S_q=e^{i\varphi_q}
\begin{pmatrix}
T_qe^{i\delta_q}&-R_qe^{-i\eta_q}\\
+R_qe^{i\eta_q}&T_qe^{-i\delta_q}
\end{pmatrix},
\label{parametrization}
\end{align}
where $T_q\equiv|T_q^\pm|$ and $R_q\equiv|R_q^\pm|$ are the transmission and the refection amplitude. The phase $\varphi_q$ is related to the determinant of the scattering matrix as $\det S_q=T_q^+/(T_q^-)^*=e^{2i\varphi_q}$. In the presence of either the time-reversal symmetry or the inversion symmetry, $\delta_q=0$.  With these definitions, the quantization condition reads
\begin{align}
\cos(qL+\varphi_{q})=T_q\cos(\theta-\delta_{q}),\quad 0<q<\pi.
\label{quantization}
\end{align}

The parametrization~\eqref{parametrization} and the quantization condition  \eqref{quantization} are invariant under the simultaneous shift $\varphi_q\to \varphi_q+\pi$, $\delta_q\to \delta_q+\pi$, and $\eta_q\to\eta_q+\pi$. To fix the ambiguity, here we assume the absence of bound states below the band bottom $\epsilon=-2t_0$.
We choose a branch of $\varphi_{q}$ in such a way that $\lim_{q\to+0}\varphi_{q}=-\pi/2$~\cite{doi:10.1063/1.531762} and is continuous as a function of $q$. See Sec.~\ref{boundstate} for the case when bound states appear below $-2t_0$.

In the defect-free case (i.e., $T_q=1$ and $\varphi_q=\delta_q=0$), the solutions to Eq.~\eqref{quantization} can be written as
\begin{align}
q_n^\pm=k_n\pm\frac{|\theta|}{L}\quad \left(k_n\equiv \frac{2\pi n}{L}\right),
\label{w0}
\end{align}
where $n=0,1,\cdots,L/2-1$ for $q_n^+$ and $n=1,2,\cdots,L/2$ for $q_n^-$. Even for a general $\hat{V}\neq0$, Eq.~\eqref{quantization} can be expressed in a form similar to Eq.~\eqref{w0}:
\begin{align}
&q_n^{\pm}=k_n+\frac{\phi_\pm(q_n^{\pm})}{L},\label{wp}\\
&\phi_\pm(q)\equiv\pm\arccos\left(T_q\cos(\theta-\delta_q)\right)-\varphi_q.\label{phipm}
\end{align}
Since the phase shift $\phi_\pm(q_n^\pm)$ in Eq.~\eqref{wp} depends on $q_n^\pm$, it still needs to be solved self-consistently.
In practice, however, one can solve it iteratively. Namely, the first approximation is to replace $q_n^\pm$ on the right hand side by $k_n$, which gives $q_n^\pm$ with an error $O(L^{-2})$. For our purpose of determining $c_{-1}(\theta)$, one needs to repeat this step once again to determine $q_n^\pm$ to the $L^{-2}$ accuracy.  We find
\begin{align}
q_n^\pm=k_n+\frac{\phi_\pm(k_n)}{L}+\frac{1}{2L^2}\frac{d[\phi_\pm(k_n)]^2}{dk_n}+O(L^{-3}).
\label{solq}
\end{align}

\subsection{Ground state energy}
\label{GSE}
Using Eq.~\eqref{solq} and performing the Taylor expansion in a series of $L^{-n}$, we evaluate the ground state energy as
\begin{align}
E_{\text{GS}}(\theta)&=-2t_0\sum_{n=0}^{\ell}\cos(q_n^+)-2t_0\sum_{n=1}^{\ell}\cos(q_n^-)\notag\\
&=\tilde{c}_{+1}^L(\theta)L+\tilde{c}_0^L(\theta)+\tilde{c}_{-1}^L(\theta)L^{-1}+O(L^{-2}).
\end{align}
Here, the coefficients $\tilde{c}_{p}^L(\theta)$ are given by
\begin{align}
\tilde{c}_{+1}^L(\theta)&\equiv-\frac{2t_0}{L}\sum_{n=-\ell}^{\ell}\cos k_n=-\frac{v_F}{\pi}-\frac{\pi v_F}{6L^2}+O(L^{-4}),\label{cpt1}
\end{align}
\begin{align}
\tilde{c}_0^L(\theta)&\equiv\frac{2t_0}{L}\sum_{n=1}^{\ell}\sin k_n\left[\phi_+(k_n)+\phi_-(k_n)\right]\notag\\
&=-\frac{2t_0}{\pi}\int_0^{k_F}dk\sin k\,\varphi_k+O(L^{-2}),
\end{align}
and
\begin{align}
\tilde{c}_{-1}^L(\theta)
&\equiv\frac{t_0}{L}\sum_{n=0}^{\ell}\frac{d}{dk_n}\left[\sin(k_n)\phi_+(k_n)^2\right]\notag\\
&\quad+\frac{t_0}{L}\sum_{n=1}^{\ell}\frac{d}{dk_n}\left[\sin(k_n)\phi_-(k_n)^2\right]\notag\\
&=\frac{v_F}{4\pi}[\phi_+(k_F)^2+\phi_-(k_F)^2]+O(L^{-1}).\label{cmt1}
\end{align}
Combining these results with Eq.~\eqref{phipm}, we find
\begin{align}
c_{+1}(\theta)&=-\frac{v_F}{\pi},\label{cp}\\
c_{0}(\theta)&=-\frac{2t_0}{\pi}\int_0^{k_F}dk\sin k\,\varphi_k,\label{c0}\\
c_{-1}(\theta)&=\frac{v_F}{2\pi}\left[\arccos\left(T_F\cos(\theta-\delta_F)\right)\right]^2+\frac{v_F}{2\pi}\varphi_{F}^2-\frac{\pi v_F}{6}.\label{cm}
\end{align}
In these expressions, $v_F$ and $k_F$ are defined in Eq.~\eqref{vF}.
In the derivation, we used the Euler--Maclaurin formula $\sum_{n=1}^{m}f(k_n)=L/(2\pi)\int_{k_1}^{k_{m+1}}dkf(k)-(1/2)[f(k_{m+1})-f(k_1)]+O(L^{-1})$ for the error estimate. The subscript $F$ refers to the value at the Fermi point $k=k_F$. The effect of $\delta_F$ merely shifts the origin of $\theta$.

We note that the present result is consistent with the earlier general observation.
The ``defect energy'' $c_0(\theta)$ is non-vanishing only in the presence of the localized defect.
On the other hand, $c_{-1}(\theta)$ is written completely in terms of the parameters of the low-energy effective theory, including the transmission amplitude $T_F$ at the Fermi point. The constant term $-\pi v_F/6$ is again the consequence of the central charge $1$ of the corresponding CFT (TLL). In this sense $c_{-1}(\theta)$ is universal.

\begin{figure}[t]
\begin{center}
\includegraphics[width=\columnwidth]{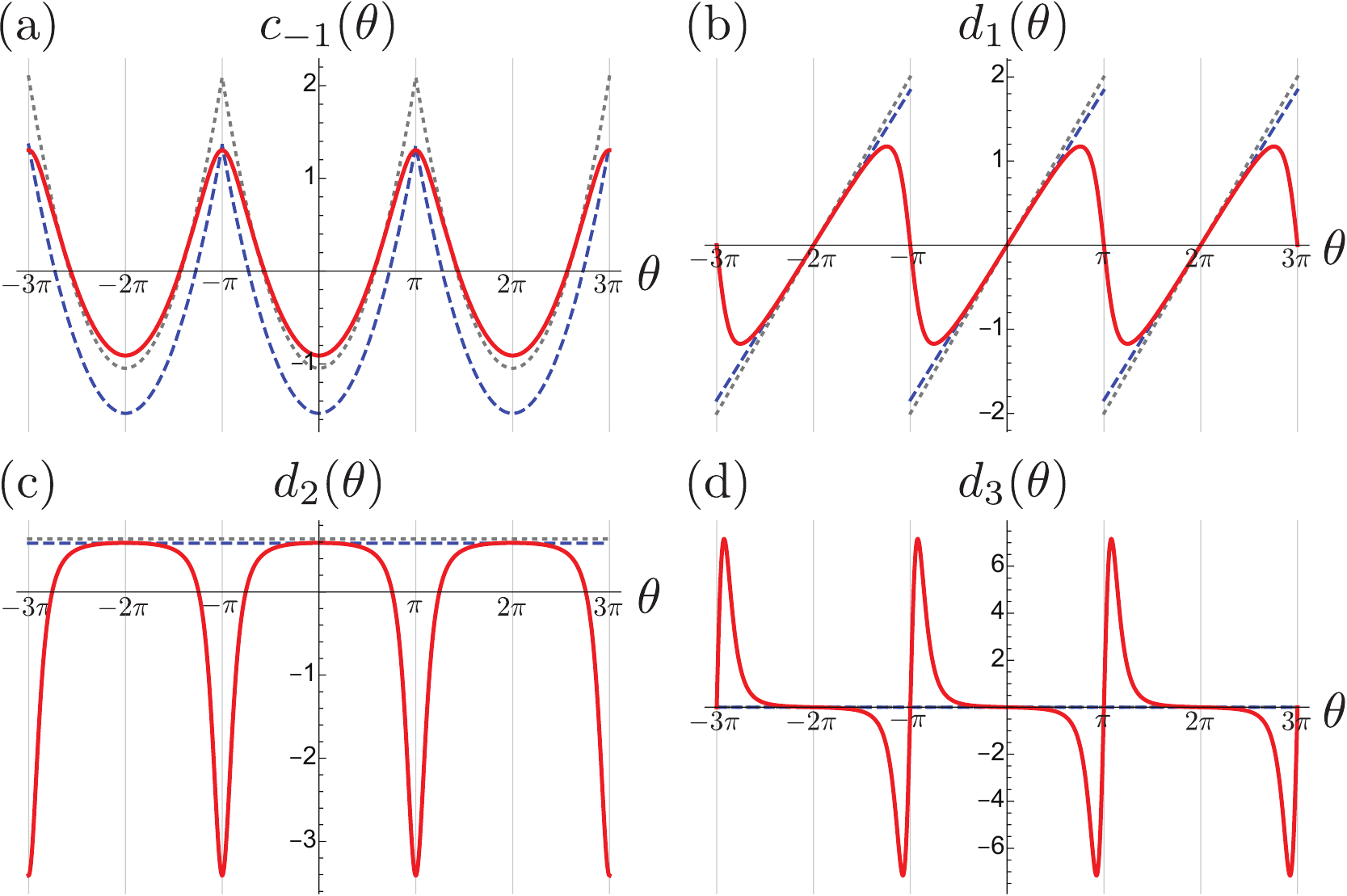}
\caption{\label{fig1}
(a) $c_{-1}(\theta)$ for the single impurity potential $w=0$ (gray dotted line) and $w=t_0=1$ (red solid line) at the half-filling. For comparison, we also plot $c_{-1}(\theta)$ for the XXZ model with $J=2$ and $\Delta=0.8$ (blue dashed line).
(b--d) The same as (a) but for $d_n(\theta)$. $n=1$ for (b), $n=2$ for (c), and $n=3$ for (d). 
}
\end{center}
\end{figure}
\subsection{Bound states}
\label{boundstate}
Some of eigenstates of $\hat{H}_0+\hat{V}$ are exponentially localized to the defect. These bound states have energy eigenvalues lower than $-2t_0$ or higher than $+2t_0$. The number of plane-wave states in Eqs.~\eqref{pw1} and \eqref{pw2} is reduced by the number of the bound states. For example, the impurity potential in Eq.~\eqref{impuritypotential} has a bound state when $|w|L> 1-\cos\theta$. The energy eigenvalue can be approximated by
$\text{sign}(w)\,\sqrt{(2t_0)^2+w^2}$ in a sufficiently large system~\footnote{More precisely, one needs to find $\lambda>0$ by solving $2t_0(\cosh L\lambda-\cos\theta)\sinh\lambda/\sinh L\lambda=|w|$. The energy eigenvalue is given by $\text{sign}(w)2t_0\cosh\lambda$.}.

Let us assume that 
there are $N_{\text{b}}$ bound states below $\epsilon=-2t_0$. We write their energy eigenvalues as $\epsilon_m^{\text{b}}$ ($m=1,2\cdots,N_{\text{b}}$). 
In this situation, we find it useful to define
\begin{equation}
\tilde{\varphi}_q\equiv\varphi_q-N_{\text{b}}\pi.
\end{equation}
The Levinson theorem states $\lim_{q\to+0}\tilde{\varphi}_{q}=-\pi/2$~\cite{doi:10.1063/1.531762}. 
Our results in Eqs.~\eqref{c0} and \eqref{cm} are modified as
\begin{align}
c_{0}(\theta)&
=-\frac{2t_0}{\pi}\int_0^{k_F}dk\sin k\,(\tilde{\varphi}_k+N_{\text{b}}\pi)+\sum_{m=1}^{N_{\text{b}}}(\epsilon_m^{\text{b}}+2t_0)\notag\\
&=-\frac{2t_0}{\pi}\int_0^{k_F}dk\sin k\,\tilde{\varphi}_k+\sum_{m=1}^{N_{\text{b}}}(\epsilon_m^{\text{b}}-\epsilon_F),\label{c0bp}\\
c_{-1}(\theta)&=\frac{v_F}{2\pi}\left[\arccos\left(T_F\cos(\theta-\delta_F)\right)\right]^2\notag\\
&\quad\quad+\frac{v_F}{2\pi}(\tilde{\varphi}_{F}+N_{\text{b}}\pi)^2-\frac{\pi v_F}{6}.\label{cmbp}
\end{align}
The first term of $c_{-1}(\theta)$, which governs the $\theta$-dependence, is independent of the bound states away from the Fermi level. The second term weakly depends on them through the choice of the branch of $\varphi_q$. This is in contrast to the non-universal defect energy $c_0(\theta)$, which explicitly depends on the bound state energy $\epsilon_m^{\text{b}}$.

\begin{figure*}[t]
\begin{center}
\includegraphics[width=17cm]{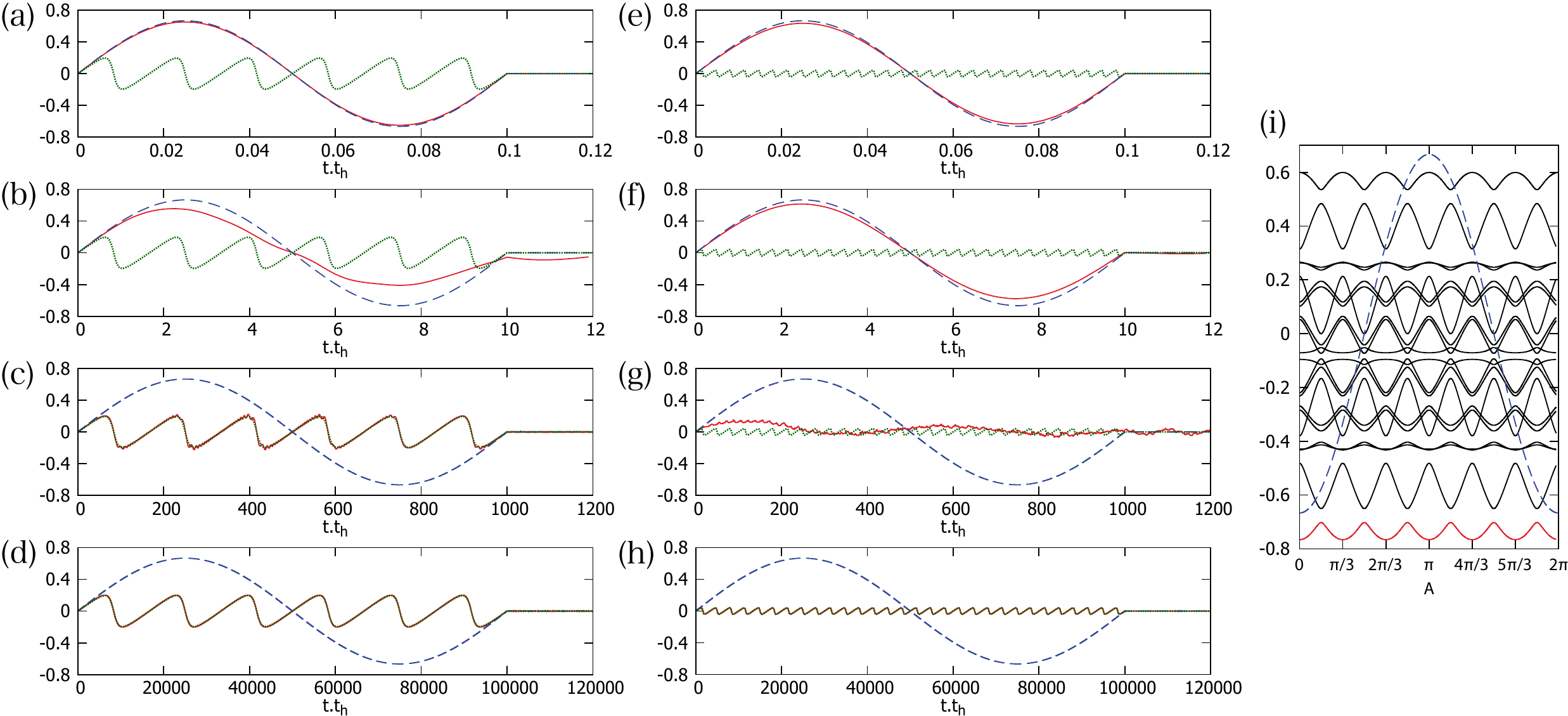}
\caption{\label{fig2}
(a--d) [(e--f)] Real-time evolution of the current density $j(t)$ (red curve) for $L=6$ [$L=30$] driven by the time-dependent Hamiltonian with a single impurity potential~\eqref{H(t)_w}. The ramp time is set to $T=10^{-1}/t_0, 10/t_0, 10^3/t_0$ and $10^5/t_0$ in (a), (b), (c) and (d) [(e), (f), (g) and (h)], respectively. The total flux $A_0$ is $2\pi$ and the defect energy $w$ is $t_0$. The blue and green dashed curve represent the adiabatic current density for $w=0$ and $w \neq 0$. (i) Many-body adiabatic spectrum of the model [Eq.~\eqref{H0}] for $w=t_0$ and $L=6$. The red curve is the ground state energy density. The blue curve denotes the sector in the adiabatic spectrum for $w=0$ connected to the ground state without flux. 
}
\end{center}
\end{figure*}

\subsection{Derivatives of ground state energy}
Given the fully general expression of $c_{-1}(\theta)$ in Eq.~\eqref{cm}, we can readily compute the derivatives of $c_{-1}(\theta)$ in Eq.~\eqref{dN}. For example,
\begin{align}
&d_1(\theta)=\frac{v_F\sin\theta'\arccos(T_F\cos\theta')}{\pi\sqrt{\sin^2\theta'+r_F^2}},
\label{d1}
\end{align}
\begin{align}
&d_2(\theta)=\frac{v_F}{\pi}\frac{\sin^2\theta'}{\sin^2\theta'+r_F^2}+d_1(\theta)\frac{r_F^2\cot\theta'}{\sin^2\theta'+r_F^2},
\label{d2}
\end{align}
and
\begin{align}
&d_3(\theta)=\frac{v_F}{2\pi}\frac{3r_F^2\sin2\theta'}{(\sin^2\theta'+r_F^2)^2}-d_1(\theta)\frac{r_F^2(2+\cos2\theta'+r_F^2)}{(\sin^2\theta'+r_F^2)^{2}},
\label{d3}
\end{align}
where $\theta'\equiv\theta-\delta_F$ and $r_F\equiv R_F/T_F$. 
Note that Eqs.~\eqref{cm} and \eqref{d1} were previously derived in Ref.~\cite{Gogolin1994} for the special case of an impurity potential in a continuum model. Here we rederived it in a  more general setting on lattice model without specifying the form of the impurity $\hat{V}$. Possible time-reversal symmetry breaking by $\hat{V}$ results in nonzero $\delta_F$ in our setting.
We plot these functions in Fig.~\ref{fig1} for the example of the single impurity potential at the half-filling, for which we have $r_F=|w|/v_F$ and $v_F=2t_0$.  We see that $d_{n}(\theta)$ generally does not vanish, implying the divergence of $\mathcal{D}_{n-1}^L(\theta)$ in the thermodynamic limit, except for several special values of $\theta$. For example, the time-reversal symmetry implies $d_{2n-1}(\theta)=0$ for $\theta=0$ and $\pi$, but $d_1(\theta),d_3(\theta)\neq0$ except for these points.  Despite this divergence, $|d_1(\theta)|$, which is the leading term in the adiabatic transport in Eq.~\eqref{jGS}, is a monotonically decreasing function of $r_F$ for a fixed $\theta$. In fact, for the present model, the frequency sum of the optical conductivity $\sigma^L(\omega)$ in Eq.~\eqref{newbound} is given by
\begin{align}
\pi L\Big\langle\frac{d^2\hat{H}}{d\theta^2}\Big\rangle=-\frac{\pi}{L}\langle\hat{H}\rangle+O(L^{-1})=v_F+O(L^{-1}),
\end{align}
implying that $|d_1(\theta)|\leq v_F$ regardless of $\theta$ or $\hat{V}$. The bound is saturated when $\theta=\pm\pi$ in the absence of the defect. Therefore, at least in this class of models, the diverging Drude weights do not imply any larger adiabatic current.

The $\theta$-dependence of $d_n(\theta)$ may be related to the singular contribution around $\theta'=\theta-\delta_F=\pi$:
\begin{align}
d_{2}(\pi+\delta_F)&=d_{2}(\delta_F)-\frac{v_F}{r_F},\label{s2}\\
d_{4}(\pi+\delta_F)&=d_{4}(\delta_F)+v_F\left(\frac{3}{r_F^3}+\frac{1}{r_F}\right),\label{s4}
\end{align}
which diverges in the $r_F=0$ limit. This divergence originates from the degeneracy of the single particle levels of $\hat{H}_0$ at $\theta=\pi$ as we show in Appendix~\ref{perturbation}. It is interesting to observe that the absolute value of the linear Drude weight can become arbitrary large without contradicting with the bound $d_2(\theta)\leq v_F/\pi$ suggested by Eq.~\eqref{drudebound}.

\section{Real-time dynamics}
\label{SecIV}
To clarify the physical implication of the divergent nonlinear Drude weights, we perform a numerical calculation. We directly simulate the real-time dynamics of the tight-binding model with a defect under a static electric field. 

We consider the dynamics driven by the time-dependent Hamiltonian, 
\begin{align}
\hat{H}(\theta(t))&=-t_0\sum_{x=1-L/2}^{L/2}(\hat{c}_{x+1}^\dagger e^{-i\theta(t)/L}\hat{c}_x+\text{h.c.})+w \hat{c}^\dagger_0 \hat{c}_0,
\label{H(t)_w}
\end{align}
where the systems size 
$L=4\ell+2$ and the number of electrons $N_{\text{el}}=L/2$. 
Here, we use a single potential disorder~\eqref{impuritypotential} as an example. Our results should be independent on the detail of the defect as discussed in Sec.~\ref{SecIII}. In Appendix~\ref{AppC}, we examine the bond disorder~\eqref{bonddisorder} and indeed obtain essentially the same result. The flux $\theta(t)$ is set to
\begin{align}
    \frac{\theta(t)}{L}=
    \begin{dcases}
    \frac{A_0}{T} t  & 0 \leq t < T,\\
    A_0 & T \leq t.
    \end{dcases} \label{theta(t)}
\end{align}
This flux insertion corresponds to the application of the static electric field $E=A_0/T$ within $0\leq t \leq T$. For convenience, we call $T$ the ramp time. The initial state $\ket{\psi_\mathrm{ini}}$ is set to the many-body ground state of $\hat{H}(0)$. For numerical calculation, we discretize the time evolution operator as $\hat{U}(t)\equiv \Delta \hat{U}_{N_t} \cdots \Delta \hat{U}_{2} \Delta \hat{U}_{1}$ where $\Delta \hat{U}_n=\exp[-i\hat{H}(t_n+\Delta t/2)\Delta t]$, $\Delta t = t/N_t$, and $t_n=(n-1)\Delta t$. $N_t$ is taken to be large enough to make the result independent on it. The observable we focus on is the time-dependent current density $j(t)$ which is defined as
\begin{align}
    j(t) = \left\langle \left.\frac{d\hat{H}(\theta)}{d\theta}\right|_{\theta=\theta(t)}\right\rangle_{\!\!t} \label{current}
\end{align}
where $\langle \cdots \rangle_t \equiv \bra{\psi(t)} \cdots \ket{\psi(t)}$ and $\ket{\psi(t)}=\hat{U}(t)\ket{\psi_\mathrm{ini}}$.

We are interested in the adiabatic transport which is dominated by the Drude weights. To clearly see it, we simulate the dynamics with changing the ramp time. There are two energy scales in this system: the energy gap induced by the defect $\Delta_V\equiv v_Fr_F/L$ and the energy scale of the flux insertion $\Delta_\theta\equiv 2\pi/T$. 
The adiabatic condition is given by $\Delta_\theta\ll\Delta_V$. This is satisfied with the sufficiently large $T$. Under this condition, the current density $j(t)$ should be governed by $j_{\text{GS}}^L(\theta)$ in Eq.~\eqref{jGS} which is directly related to the Drude weights. On the other hand, $\Delta_\theta\gg\Delta_w$ corresponds to the sudden quench.

The current density $j(t)$ is shown in Fig.~\ref{fig2}~(a)-(h). Figs.~2~(a)~and~(e) are near the quench limit. They show the usual Bloch oscillation with the period $T_B=2\pi/E=(2\pi/A_0)T$. The behavior is almost the same as the $r_F=0$ case which is $j(t)=(2 t_0/\pi)\sin(\theta(t)/L)$ denoted by the blue dashed curve. Making the ramp time longer, the current amplitude is suppressed [Figs.~2~(b)~and~(f)] and the profile approaches to the different oscillational modes [Figs.~2~(c)~and~(g)]. Finally, it reaches the different periodic oscillation profile which has a smaller amplitude and shorter period $T_B^\prime=T_B/L$ [Figs.~2~(d)~and~(h)]. This profile reaches the leading term of the adiabatic current with the defect, i.e., $j(t) \sim d_1(\theta(t))/L$. Therefore, this dynamics reaches the adiabatic limit. As discussed above, the tight-binding model with a defect shows the divergent nonlinear Drude weights. However, the current response is not enhanced with increasing the system size. This shows that the divergence does not necessarily imply the large current response. This observation is consistent with the argument from the analytical results in Sec.~III~E. 

The result of our real-time simulation suggests a way to experimentally measure the nonlinear Drude weights through an experiment on the persistent current.
For example, both the flux insertion and the transport measurement have been realized in ultracold atomic systems~\cite{Amico2005, Ryu2007, Cominotti2014, Eckel2014, Lacki2016}. The mesoscopic systems like a metallic ring can also be a realistic platform where the persistent current has been accurately measured~\cite{Bleszynski-Jayich2009, Bluhm2009}.

Finally, we mention that the crossover-like behavior with changing the ramp time is understood from the many-body adiabatic spectrum shown in Fig.~\ref{fig2}~(i). In the quench limit, the defect energy scale $\Delta_V$ is much smaller than the flux insertion one $\Delta_\theta$ and thus the defect should be irrelevant. Without the defect, the momentum of each electron is conserved and the adiabatic spectrum is $2\pi$-periodic in therms of $\theta/L$ [the blue dashed curve in Fig.~\ref{fig2}~(i)]. This periodicity appears as the Bloch oscillation in the quench limit. Even with the defect, almost perfect non-adiabatic transitions occur at every gaps induced by the defect and then the $2\pi$-periodic behavior appears as shown in Fig.~\ref{fig2}~(a)~and~(e). On the other side, there does not happen such non-adiabatic transitions and the time-dependent state keeps sitting on the instantaneous ground state. Such trajectory is shown in the red curve in Fig.~\ref{fig2}~(i) and this corresponds to the small oscillation with the period $T_B^\prime$ in the adiabatic limit [Figs.~\ref{fig2}~(d)~and~(h)]. In the intermediate scale, non-adiabatic transitions occur in a very complex way and there also appear interference in the spectrum. Note that a similar crossover behavior and the interference effect appears in the flux insertion in the 1D Hubbard model~\cite{Oka2003, Oka2005QW} and this kind of behavior is expected to appear typically in the many-body quantum systems.

\section{Relation to  boundary confromal field theory}
\label{SecV}
In this section, we mention the relation between our results and the boundary CFT. In fact, the first term of $c_{-1}(\theta)$ in Eq.~\eqref{cm} which depends on $T_F$ can be interpreted in terms of the boundary CFT. 

The defect in the tight-binding model we consider corresponds to the barrier in a TLL studied in Refs.~\cite{KaneFisher-PRL1992,KaneFisher-PRB1992} at the free fermion point $K=1$.
The system studied in this paper is a finite ring of circumference $L$ with the single defect.
Such a system can be mapped to a problem of boundary CFT by a folding trick~\cite{WongAffleck,IsingDefect-PRL,Yjunction-JSTAT}: after the folding, we have a 
two-component TLL of length $l=L/2$ with two boundaries: one corresponding to the defect and the other corresponding to no defect.
Although the problem at this point is a two-component TLL (of central charge 2) with boundaries, we can decompose the two-component TLL into
even and odd combinations of the original fields $\phi(x) \pm \phi(-x)$.
It can be seen that the odd component does not ``feel'' the defect and is always subject to the same boundary condition.
Thus the problem is effectively reduced to the single-component TLL (of the even field) with boundaries, although a care should be taken about ``gluing condition''~\cite{WongAffleck,Yjunction-JSTAT,Oshikawa2010BCFT} in reconstructing the spectrum of the original model.
In the discussion of the universal part of the ground-state energy, which is the focus of the present paper, however, the gluing condition is not important and we can simply study the single-component TLL with boundaries.
For generic values of the Luttinger parameter $K$, the barrier is either a relevant or irrelevant perturbation (in the renormalization group sense), so that the defect is renormalized into an infinitely strong barrier which completely reflects the current, or a vanishing barrier which transmits the current perfectly.
In terms of the (even component of) ``phase field'' of the TLL, the infinitely strong barrier corresponds to the Neumann boundary condition, while the vanishing barrier corresponds to the Dirichlet boundary condition.
The free fermion case $K=1$ corresponds to the boundary between the two phases.
Here the barrier is an exactly marginal perturbation, so that there is a continuous family of the boundary conditions interpolating the vanishing barrier and the infinitely strong barrier. This exactly corresponds to the $S$-matrix of the defect for the incoming free fermions.

In fact, the continuous family of the boundary conditions at $K=1$ was studied in terms of free bosons in Refs.~\cite{CallanKlebanov,CKLM1994} and in terms of free fermions in Ref.~\cite{PolchinskiThorlacius}.
Here we show that our result agrees with theirs.
In Refs.~\cite{CallanKlebanov,CKLM1994}, the continuous family of the boundary conditions at $K=1$ is formulated in terms of the emergent SU(2) symmetry.
This SU(2) degree of freedom indeed corresponds to the $S$-matrix of free fermions.
In Ref.~\cite{PolchinskiThorlacius}, the $S$-matrix was parametrized in terms of the effective barrier parameter $g$ (complex number).
For simplicity here we focus on the case $g$ is real, for which the $S$-matrix is given as $S \sim \exp{(i \pi g \sigma_1)}$ and thus $T^\pm_F = T_F = \cos{(\pi g)}$.
The phase shift in those papers, for example in Eq.~(13) of Ref.~\cite{PolchinskiThorlacius}, then simply reads $\Delta = \pi |g| = \arccos{T_F}$.

The partition function for the boundary condition corresponding to the the barrier strength $g$ on one side and the Neumann boundary condition (corresponding to the infinite barrier) on the other side (in the even fermion number sector) at the inverse temperature $\beta$ is~\cite{PolchinskiThorlacius}
\begin{equation}
    Z_{BN}(q) \sim \frac{q^{-1/24}}{\prod_{m=1}^\infty (1-q^m)} \sum_{k=0}^\infty q^{\frac{1}{2} \left[ k + \frac{1}{2} - (-1)^k \frac{\Delta}{\pi} \right]}, 
\end{equation}
where $q \equiv e^{-\pi \beta/l}$.
In our setup, we just consider a single barrier, so that the other side obeys the Dirichlet boundary condition after the folding.
In this case, the partition function is rather given by
\begin{equation}
    Z_{BD}(q) \sim \frac{q^{-1/24}}{\prod_{m=1}^\infty (1-q^m)} \sum_{k=0}^\infty q^{\frac{1}{2} \left[ k +  (-1)^k \frac{\Delta}{\pi} \right]} .
\end{equation}
We can read off the ground-state energy (relative to $g=0$ case) as
\begin{equation}
    \frac{\pi}{l} \frac{1}{2} \left( \frac{\Delta}{\pi} \right)^2 = \frac{1}{L} \frac{\Delta^2}{\pi} . 
\end{equation}
Because $\Delta = \arccos{T_F}$, this agrees with our Eq.~\eqref{cm} with $\delta_F=\theta=\varphi_F=0$ and $v_F=1$.

\section{Summary and Outlook}
\label{SecVI}
In this work, we clarified the dependence of the ground state energy on the twisted boundary condition in 1D systems in general. As an illuminating toy model, we discussed a single-band tight-binding model in the presence of a single defect. We found that the linear Drude weight $\mathcal{D}_1^L(\theta)$ in the large $L$ limit depends non-trivially on the twist angle $\theta$ due to the presence of the defect. As stressed in the introduction, our viewpoint is different from the previous studies~\cite{RigolShastry-Drude,BellomiaResta} that discussed how to extract the Drude weight under open boundary conditions. We also found that $N$-th order Drude weights $\mathcal{D}_N^L(\theta)$ ($N\geq2$) in our model exhibits a strong divergence proportional to $L^{N-1}$ in the large $L$ limit. In addition, we studied the physical implication of the divergence through the direct numerical simulation which is relatively easy since this model is non-interacting. This model will help us further develop general theories of transport properties with or without the translation invariance.

There remains various open issues for future work. While we only studied a single defect, it is an attractive problem to study the multi-defect (disordered) model, which should be relevant to the physics of Anderson localization. The effect of interaction in the single-defect model is also interesting because it has been already known that the transport properties of this model can be drastically changed by an interaction~\cite{KaneFisher-PRL1992, KaneFisher-PRB1992, Loss1992, Furusaki-Nagaosa1993, Gogolin1994, Meden2003, Dias2006}. 
It should be clarified how the pathological behaviors for the nonlinear Drude weights presented in this paper is modified by the interaction.

It is also important to further develop the general theory for the adiabatic transport. 
Our discussion in Sec.~\ref{general} suggests that  the defect energy $c_0(\theta)$ vanishes and that $c_{-1}(\theta)$ is a quadratic function of $\theta$, i.e., $d_n(\theta)=0$ ($n\geq3$) in any systems with a U(1) symmetry and a lattice translation symmetry, regardless of whether the low-energy effective theory of the system is TLL or not. In particular, $d_n(\theta)=0$ ($n\geq3$) is the condition for the linear Drude weight to be $\theta$-independent in the large $L$ limit. These statements may be rationalized by the following argument. Let us imagine dividing the system into $M$ subsystems ($M\geq2$) in such a way that each part has the length $L_i\gg1$ and $\sum_{i=1}^ML_i=L$. We decompose $\theta$ correspondingly into $\theta_i\equiv\theta L_i/L$ ($i=1,\cdots,M$). We expect that the ground state energy of the $i$-th part is given by $E_{\text{GS}}^{L_i}(\theta_i)=\sum_{p=+1,0,-1,\cdots}c_p(\theta_i)L_i^p$ and that the total ground state energy satisfies the additivity $\sum_{i=1}^ME_{\text{GS}}^{L_i}(\theta_i)=E_{\text{GS}}^{L}(\theta)+O(L^{-1})$. This is possible only when $c_{0}(\theta)$, which should be independent of $\theta$, vanishes. Furthermore, we demand that the leading $L^{-1}$ correction is $\theta$-independent, i.e., $\sum_{i=1}^M[E_{\text{GS}}^{L_i}(\theta_i)-E_{\text{GS}}^{L_i}(0)]=[E_{\text{GS}}^{L}(\theta)-E_{\text{GS}}^{L}(0)]+o(L^{-1})$, which suggests that $c_{-1}(\theta)-c_{-1}(0)$ is proportional to $\theta^2$. We leave more rigorous proof of these conjectures to future work.

\begin{acknowledgments}
We thank Yohei Fuji, Hideaki Obuse, Yuhi Tanikawa, Hosho Katsura, Wonjun Lee, and Gil-Young Cho for valuable discussions. The work of K.T. is supported by the U.S. Department of Energy (DOE), Office of Science, Basic Energy Sciences (BES), under Contract No. AC02-05CH11231 within the Ultrafast Materials Science Program (KC2203). K. T. also thanks JSPS for support from Overseas Research Fellowship. The work of M. O. is supported in part by MEXT/JSPS KAKENHI Grant Nos.~JP17H06462 and JP19H01808, JST CREST Grant No. JPMJCR19T2. The work of H. W. is supported by JSPS KAKENHI Grant No.~JP20H01825 and by JST PRESTO Grant No.~JPMJPR18LA. 
\end{acknowledgments}

\appendix

\section{Optical conductivity}
\label{optical}
In Kubo's theory, the optical conductivity $\sigma^L(\omega)$ is given by~\cite{Resta_2018,WatanabeLiuOshiakwa}
\begin{equation}
\sigma^L(\omega)=\frac{i}{\omega+i\eta}\phi^L(\omega),
\end{equation}
where
\begin{align}
\phi^L(\omega)&\equiv L\Big\langle \frac{d^2\hat{H}}{d\theta^2}\Big\rangle\notag\\
&\quad+L\Big\langle\frac{d\hat{H}}{d\theta}\frac{\hat{Q}}{\omega-(\hat{H}-E_{\text{GS}}^L)+i\eta}\frac{d\hat{H}}{d\theta}\Big\rangle\notag\\
&\quad-L\Big\langle\frac{d\hat{H}}{d\theta}\frac{\hat{Q}}{\omega+(\hat{H}-E_{\text{GS}}^L)+i\eta}\frac{d\hat{H}}{d\theta}\Big\rangle
\end{align}
and $\hat{Q}$ is the projection operator onto excited states. The expectation value is taken using the ground state. 
The real part of the conductivity may be written as
\begin{align}
\label{resigma}
\text{Re}[\sigma^L(\omega)]=\delta(\omega)\pi\mathcal{D}_1^L+\text{Re}[\sigma_{\text{reg.}}^L(\omega)],
\end{align}
where
\begin{align}
\mathcal{D}_1^L= L\Big\langle \frac{d^2\hat{H}}{d\theta^2}\Big\rangle-2L\Big\langle\frac{d\hat{H}}{d\theta}\frac{\hat{Q}}{\hat{H}-E_{\text{GS}}^L}\frac{d\hat{H}}{d\theta}\Big\rangle=L\frac{d^2E_{\text{GS}}^L(\theta)}{d\theta^2}
\end{align}
is the (linear) Drude weight and 
\begin{align}
\text{Re}[\sigma_{\text{reg.}}^L(\omega)]&=\pi L\Big\langle\frac{d\hat{H}}{d\theta}\frac{\hat{Q}\delta(|\omega|-(\hat{H}-E_{\text{GS}}^L))}{\hat{H}-E_{\text{GS}}^L}\frac{d\hat{H}}{d\theta}\Big\rangle\geq0
\end{align}
is the real part of the regular part of $\sigma^L(\omega)$.

The frequency sum rule reads
\begin{equation}
\int_{-\infty}^{\infty}d\omega \sigma^L(\omega)=\int_{-\infty}^{\infty}d\omega \text{Re}[\sigma^L(\omega)]=\pi L\Big\langle\frac{d^2\hat{H}}{d\theta^2}\Big\rangle.
\end{equation}
Since the second and the third term of Eq.~\eqref{resigma} are non-negative, the frequency sum must cannot be smaller than $\pi\mathcal{D}_1^L$. This justifies Eq.~\eqref{drudebound} in the main text.

\begin{figure*}[t]
\begin{center}
\includegraphics[width=14cm]{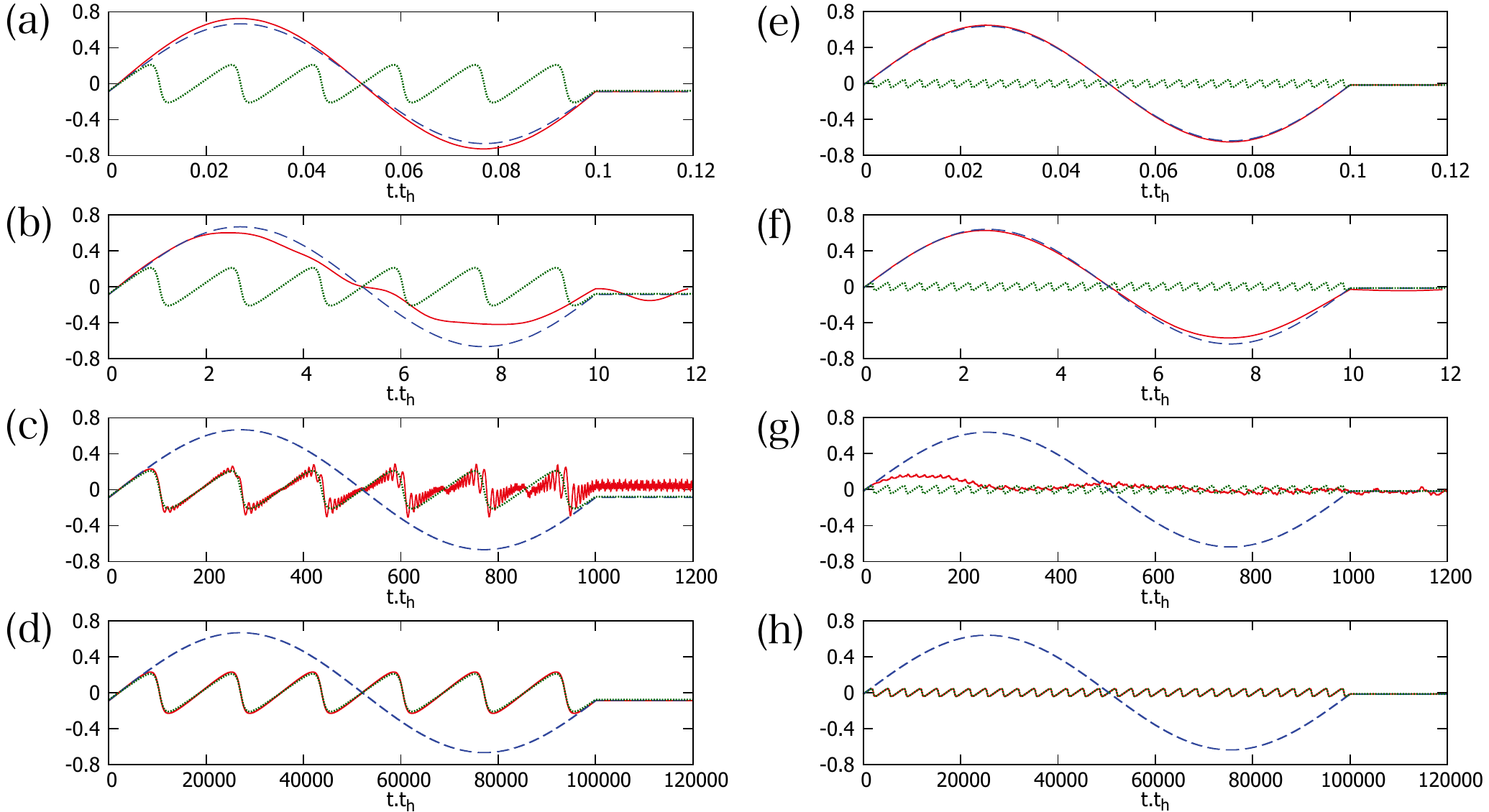}
\caption{\label{fig3}
(a--d) [(e--f)] Real-time evolution of the current density $j(t)$ (red curve) for $L=6$ [$L=30$] driven by the time-dependent Hamiltonian with a bond disorder~\eqref{H(t)_v}. The ramp time is set to $T=10^{-1}/t_0, 10/t_0, 10^3/t_0$ and $10^5/t_0$ in (a), (b), (c) and (d) [(e), (f), (g) and (h)], respectively. The total flux $A_0$ is $2\pi$. The parameters for the disorder are set to $v=1.5t_0$ and $\delta=\pi/4$. The blue and green dashed curve represent the adiabatic current density for $v=0$ and $v \neq 0$.
}
\end{center}
\end{figure*}

\section{Perturbative justification of the main result}
\label{perturbation}
Some key aspects of $c_{-1}(\theta)$ derived in Sec.~\ref{GSE} of the main text can be readily understood by treating $\hat{V}$ as perturbation. Here we use the single impurity potential~\eqref{impuritypotential} as an example. As noted before, $r_F\equiv R_F/T_F=|w|/v_F$ and $\delta_F=0$ in this case. The matrix element $\langle k_n|\hat{V}| k_n\rangle=w/L$ is independent of $n$ and $m$ and is inversely proportional to the system size. 

The first-order correction to the ground-state energy is
\begin{equation}
\sum_{n:\text{occ}.}\langle k_n|\hat{V}| k_n\rangle=w\frac{N_{\text{el}}}{L}
\end{equation}
This is a part of the defect energy $c_0(\theta)$.

The second-order correction is given by
\begin{align}
&-\sum_{n:\text{occ.}}\sum_{m:\text{unocc.}}\frac{\langle k_n|\hat{V}| k_m\rangle\langle k_m|\hat{V}|k_n\rangle}{\epsilon_{k_m}-\epsilon_{k_n}}\notag\\
&=-\frac{v_F^2r_F^2}{L^2}\sum_{n:\text{occ.}}\sum_{m:\text{unocc.}}\frac{1}{\epsilon_{k_m}-\epsilon_{k_n}}
\label{E2}
\end{align}
To extract the most singular contributions from adjacent of the Fermi points $k=\pm k_F$, we linearize the dispersion as $\epsilon_{k}=\pm v_F(k \mp k_F+\theta/L)$.  The summation in Eq.~\eqref{E2} is dominated by the scattering between $k=k_F$ and $-k_F$:
\begin{align}
&-\frac{v_Fr_F^2}{L^2}\sum_{k_n>-k_F}\sum_{k_m>k_F}\frac{1}{k_m+k_n+2\theta/L}\notag\\
&\quad\quad+\frac{v_Fr_F^2}{L^2}\sum_{k_n<k_F}\sum_{k_m<-k_F}\frac{1}{k_m+k_n+2\theta/L}\notag\\
&=-\frac{v_Fr_F^2}{2\pi L}\sum_{n,m=0}^\infty\frac{2(m+n+1)}{(m+n+1)^2-(\theta/\pi)^2}\notag\\
&=\frac{v_Fr_F^2}{2\pi L}\theta\cot \theta+\dots.
\end{align}
In the last step, we regularize the summation by subtracting the $\theta$-independent term.  This gives the correct $\theta$-dependence of $c_{-1}(\theta)$ in Eq.~\eqref{cm} up to $r_F^2$, and Eqs.~\eqref{d1}--\eqref{d3} can be fully reproduced up to this order of $r_F$.

The above perturbation theory fails near $\theta=\pi$ because of the degeneracy of the $N_{\text{el}}$-th level and the $(N_{\text{el}}+1)$-th level ($n=\ell$ and $n=-\ell-1$) of $\hat{H}_0$.
Focusing only on these two levels, we find
\begin{equation}
\begin{pmatrix}
\epsilon_{k_{\ell}}&0\\
0&\epsilon_{k_{-\ell-1}}
\end{pmatrix}
+\frac{w}{L}
\begin{pmatrix}
1&1\\
1&1
\end{pmatrix}.
\end{equation}
The smaller eigenvalue of this matrix is
\begin{align}
&\frac{w}{L}+\varepsilon_F\cos\left(\frac{\theta-\pi}{L}\right)-\sqrt{v_F^2\sin^2\left(\frac{\theta-\pi}{L}\right)+\frac{w^2}{L^2}}, 
\end{align}
implying that $c_{-1}(\theta)$ contains
\begin{align}
&-v_F\sqrt{(\theta-\pi)^2+r_F^2}\notag\\
&=-v_F\sum_{N=0}^\infty \frac{\Gamma(\tfrac{3}{2})}{N!\Gamma(\tfrac{3}{2}-N)}(\theta-\pi)^{2N}r_F^{1-2N}.
\end{align}
Hence, the most singular term in $d_{2N}(\theta=\pi)$ ($N\geq 1$) in the $r_F=+0$ limit is given by 
\begin{align}
-v_F\frac{(2N)!\Gamma(\tfrac{3}{2})}{N!\Gamma(\tfrac{3}{2}-N)}\frac{1}{r_F^{2N-1}}.
\end{align}
This reproduces our results in Eqs.~\eqref{s2} and \eqref{s4}.

\section{Real-time simulation for a bond disorder}
\label{AppC}
Our main claim in this paper does not depend on the detail of the defect. This is because the universal dependence of the ground state energy on the twist angle is fully characterized by the transmission coefficient of the defect scattering as shown in Sec.~\ref{SecIII}. To support this point from the numerical calculation, here we examine the bond disorder in~\eqref{bonddisorder} as our second example.

We apply a static electric field to the tight-binding model with the bond disorder. The time-dependent Hamiltonian is given by
\begin{align}
\hat{H}(\theta(t))&=-t_0\sum_{x=1-L/2}^{L/2}(\hat{c}_{x+1}^\dagger e^{-i\theta(t)/L}\hat{c}_x+\text{h.c.}) \nonumber\\
&\qquad+ \{-(ve^{i\delta}-t_0)\hat{c}_{1}^\dagger e^{-i\theta(t)/L}\hat{c}_{0}+\text{h.c.}\},
\label{H(t)_v}
\end{align}
where the systems size 
$L=4\ell+2$ and the number of electrons $N_{\text{el}}=L/2$. The time dependence of the flux $\theta(t)$ is given by Eq.~\eqref{theta(t)}. The time-evolution protocol and the calculation method are the same as in Sec.~\ref{SecIV}.

We calculate the current density $j(t)$ defined by Eq.~\eqref{current} and the results are shown in Fig.~\ref{fig3}. The qualitative behavior is the same as the potential disorder case shown in~Fig.~\ref{fig2}. This supports the validity of our analytical results in Sec.~\ref{SecIII}. Also, this result suggests that the current response is weakened by the defect while the defect induces the divergence of nonlinear Drude weights. A slight difference from the potential disorder is the time-reversal symmetry breaking effect, i.e., $\delta \neq 0$. This makes the persistent current nonvanishing even without an electric field.

\bibliography{ref.bib}

\clearpage

\end{document}